# Affine subgroups of the affine Coxeter group with the same Coxeter number

Nazife Ozdes Koca[1,*] and Mehmet Koca[2]

[1] Department of Physics, College of Science, Sultan Qaboos University, P.O. Box 36, Al-Khoud 123, Muscat, Sultanate of Oman
[2] Professor Emeritus, Department of Physics, College of Science, Sultan Qaboos University, P.O. Box 36, Al-Khoud 123, Muscat, Sultanate of Oman
*Corresponding author email address: nazife@squ.edu.om

## ABSTRACT

Affine subgroups having the same Coxeter number with the affine Coxeter groups $W(A_n)$, $W(D_n)$ and $W(E_n)$ are constructed by graph folding technique. The affine groups $W(C_n)$ and $W(B_n)$ are obtained from the Coxeter groups $W(A_{2n-1})$ and $W(D_{2n-2})$ respectively. The affine groups $W(E_6)$, $W(D_6)$ and $W(E_8)$ lead to the affine groups $W(F_4), W(H_3)$ and $W(H_4)$ respectively by graph folding. The latter two are the non-crystallographic groups where $W(H_3)$ plays a special role in the quasicrystallographic structures with icosahedral symmetry. A general construction of the affine dihedral subgroups is introduced, some of which, describe the existing planar quasicrystallography. In the construction of the root systems, sets of orthonormal vectors are used but a special non-orthogonal set of vectors in the formulation of the root system of $W(A_n)$ is also introduced which has practical applications in the construction of the lattices $A_n$ and $A_n^*$ and their Delone and Voronoi cells.

## 1. Introduction

To set the scene we introduce the notation adopted by Conway & Sloane [1-2] where the lowercase letters $a_n, b_n, c_n, d_n, e_n$ ($n = 6, 7, 8$), $f_4$ and $g_2$ represent the Coxeter-Dynkin diagrams. The corresponding point groups with their orders and the Coxeter numbers $h$ given in parathesis are as follows: $W(a_n)((n+1)!, h = n+1)$, $W(b_n)(2^n n!, h = 2n)$, $W(c_n)(2^n n!, h = 2n)$, $W(d_n)(2^{n-1}n!, h = 2n-2)$, $W(e_n)(n = 6, 7, 8), W(f_4)$ and $W(g_2)$ for the Coxeter-Weyl groups generated by reflections. The orders and the Coxeter numbers of the latter point groups are $(2^7\ 3^4\ 5, h = 12)$ for $W(e_6)$, $(2^{10}\ 3^4\ 5\ 7, h = 18)$ for $W(e_7)$, $(2^{14}\ 3^5\ 5^2\ 7, h = 30)$ for $W(e_8)$, $(1152, h = 12)$ for $W(f_4)$ and $(12,, h = 6)$ for $W(g_2)$. The capital letters $A_n, B_n, C_n, D_n, E_n (n = 6, 7, 8)$, $F_4$ and $G_2$ represent the extended Coxeter-Dynkin diagrams and the corresponding infinite groups $W(A_n), W(B_n), W(C_n), W(D_n), W(E_n)(n = 6, 7, 8), W(F_4)$ and $W(G_2)$ are the affine groups generated by the reflections defined by the extended diagrams. The root lattices $\Lambda$ are denoted by $A_n$ for $W(A_n)$, $D_n$ for $W(D_n)$, $E_n$ for $W(E_n)$ and $\mathbb{Z}^n$ for $W(B_n)$ and $D_n$ for $W(C_n)$, $D_4$ for $W(F_4)$ and $A_2$ for $W(G_2)$. The weight lattices $\Lambda^*$ are obtained by starring the root lattices. However, some of the lattices are self-dual like $A_2, D_4, \mathbb{Z}^n$ and $E_8$. The diagrams corresponding both to the point groups as well as the infinite simply-laced groups are shown in Figure 1. Relations of the Coxeter-Weyl groups to the Lie algebras can be found in the reference [3].

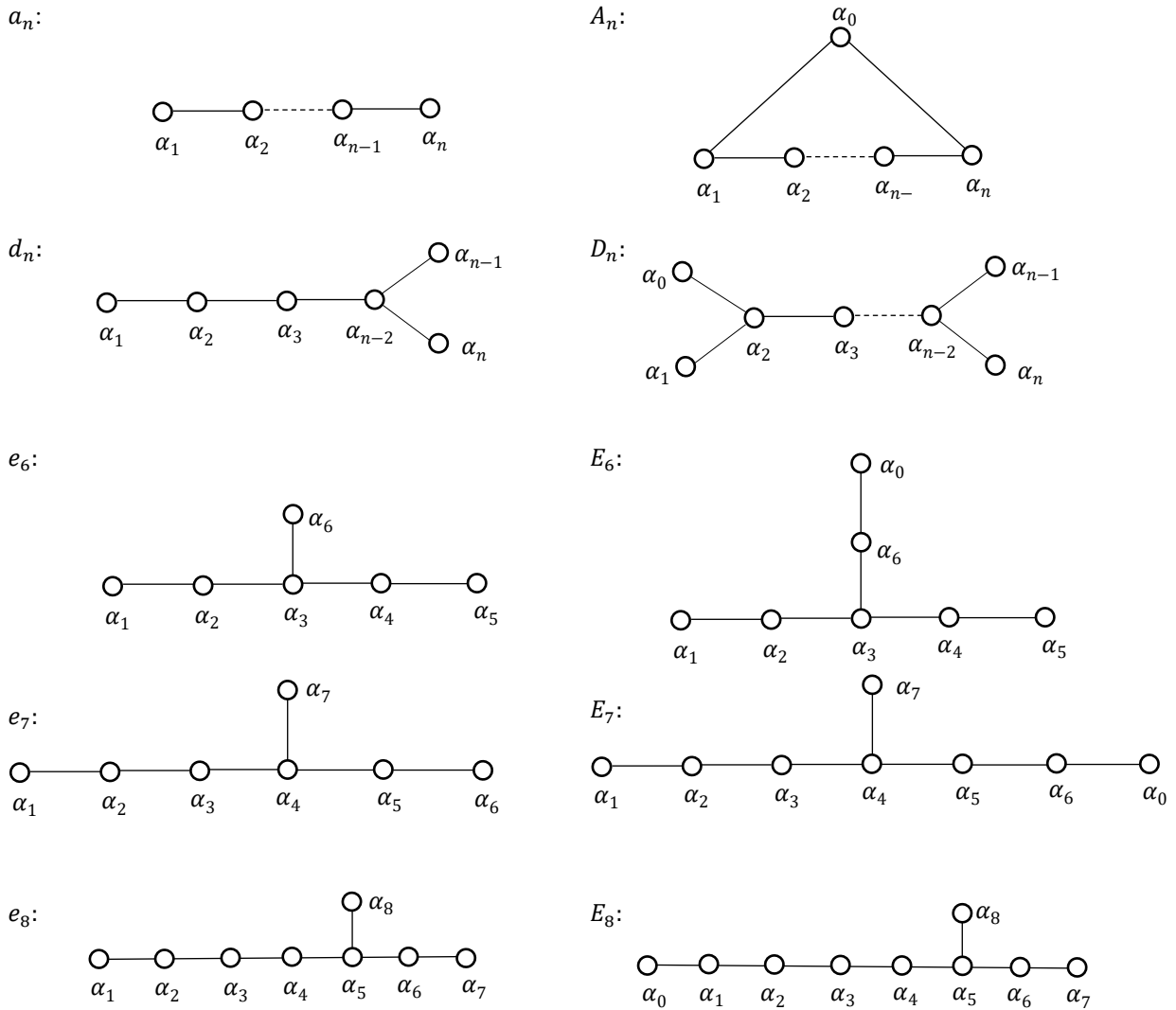

**Figure 1**. Coxeter-Dynkin diagrams for the simply-laced finite and infinite groups.

The affine reflection of a vector $\lambda$ for an arbitrary root $\alpha$ and the integer $n$ is defined as

$$r_{\alpha,n}(\lambda) = \lambda - \frac{2((\lambda,\alpha)-n)\alpha}{(\alpha,\alpha)}. \tag{1}$$

We introduce the set of orthonormal vectors $l_i$ $(i = 1, 2, \ldots, n)$, $(l_i, l_j) = \delta_{ij}$ by which we can construct the simple roots of the corresponding Lie algebras and the lattice vectors can be expressed in terms of these orthonormal set of vectors. However, for the Coxeter group $W(a_n)$ we introduce a new set of vectors $k_i$ $(i = 1, 2, \ldots, n+1)$ defined by $k_i = l_i - \frac{l_0}{n+1}$, with $l_0 = \sum_{i=1}^{n+1} l_i$ satisfying the relations $\sum_{i=1}^{n+1} k_i = 0$ and the inner products $(k_i, k_i) = \frac{n}{n+1}$ and $(k_i, k_j) = -\frac{1}{n+1}$ for $i \neq j$. These vectors are useful as they represent the vertices of the simplices of the groups $W(a_n)$. In the rest of the work we will construct the subgroups of the groups $W(A_n), W(D_n), W(E_n)$ sharing the same Coxeter number [4] with the parent group.

## 2. The Group $W(A_n)$

The simple roots of the point group $W(a_n)$ can be defined as

$$\alpha_i = k_i - k_{i+1}, \quad i = 1, \ldots, n, \quad \alpha_0 = k_{n+1} - k_1 \tag{2}$$

and the corresponding weights defined by $\left(\omega_i, \frac{2\alpha_j}{(\alpha_j,\alpha_j)}\right) = \delta_{ij}$ can be obtained as

$$\omega_i = k_1 + k_2 + \cdots k_i. \tag{3}$$

The point group $W(a_n)$ acts as permutations on the vectors $k_i$. The root lattice $A_n$ consists of the vectors as the integer linear combinations of the simple roots. The polytope consisting of the vertices of the set of vectors $(k_i - k_j)$, $i \neq j = 1, \ldots, n+1$ is called the contact polytope (root polytope) of the root lattice. The orbits of the weights under the point group $W(a_n)\omega_i$ represent the vertices of the ambo-simplexes of all orders. For instance, vertices of the simplex $W(a_n)\omega_1$ consists of the vectors $k_1, k_2, \ldots, k_{n+1}$ and similarly the simplex $W(a_n)\omega_n$ consists of the vertices $-k_1, -k_2, \ldots, -k_{n+1}$.

The fundamental simplex consists of the $n+1$ vertices $0, \omega_1, \omega_2, \ldots, \omega_n$. The Voronoi polytope $V(0)$ is the dual of the root polytope which consists of the vertices of the union of the orbits of the weights under the point group $W(a_n)\omega_i$ [5]. The Delone polytopes nearest to the vertices of the Voronoi cell $V(0)$ consist of the lattice points as the sum of the paired simplexes such as

$$W(a_n)\omega_1 + W(a_n)\omega_n, W(a_n)\omega_2 + W(a_n)\omega_{n-1}, \ldots. \tag{4}$$

The dual lattice $A_n^*$ of the lattice $A_n$ consists of the vectors as the integer linear combinations of the fundamental weights $\omega_i$. That is to say the weight lattice $A_n^*$ consists of the images of all vertices of the fundamental simplex under the group $W(A_n)$. The orbit of the centroid of the fundamental simplex,

$$P = \frac{0+\omega_1+\cdots+\omega_n}{n+1} \tag{5}$$

can be written as

$$P = \frac{(n+1)\,k_1+nk_2+\cdots+k_{n+1}}{n+1}. \tag{6}$$

Since the point group $W(a_n)$ permutes the vectors $k_i$ the set of vectors $(n+1)W(a_n)P$ is defined as the permutohedron of rank $n$ with $(n+1)!$ vertices. The Delone polytope is the fundamental simplex centered at $P$ and the Voronoi polytope is the permutohedron [6]. The contact polytope has the vertices $\pm k_i$, $(i = 1, 2, ..., n+1)$ known as the diplo-simplex.

We will discuss the group $W(C_n)$ as the subgroup of the group $W(A_{2n-1})$ since both have the same Coxeter number $h = 2n$. We will also discuss the dihedral point subgroup $W(I_2(h))$ of the group $W(a_n)$ and its affine extension. Dihedral subgroup of the group $W(a_n)$ is important for the lattice projection onto the Coxeter plane for they may serve for the quasicrystallography [7]. For this purpose, a useful coordinate system can be determined for the vectors $k_j$ as

$$k_j = \sqrt{\frac{2}{h}} \left( e^{i\frac{2\pi j}{h}}, e^{i\frac{4\pi j}{h}}, ..., e^{i\frac{(h-2)\pi j}{h}}, \frac{1}{\sqrt{2}} \right), (j = 1, 2, ..., n+1), \text{ for } n \text{ odd.} \quad (7)$$

$$k_j = \sqrt{\frac{2}{h}} \left( e^{i\frac{2\pi j}{h}}, e^{i\frac{4\pi j}{h}}, ..., e^{i\frac{(h-2)\pi j}{h}}, e^{i\frac{(h-1)\pi j}{h}} \right), (j = 1, 2, ..., n+1), \text{ for } n \text{ even.} \quad (8)$$

They are useful for the lattice projection onto the Coxeter plane with the dihedral symmetry as the first complex component $e^{i\frac{2\pi j}{h}}$ of $k_j$ is sufficient.

## 2.1 Affine Dihedral Subgroup of the affine Coxeter group $W(A_n)$

Generators of the dihedral subgroup [8] of the point group $W(a_n)$ can be defined by $R_1 = r_{\alpha_1} r_{\alpha_3} ...$, $R_2 = r_{\alpha_2} r_{\alpha_4} ...$, as the products of the reflection generators with respect to the simple roots partitioned as sets of orthogonal vectors. $R_1$ and $R_2$ are the generators of the dihedral subgroup $W(I_2(h))$ of the point Coxeter group satisfying the relations $R_1^2 = R_2^2 = (R_1 R_2)^h = 1$. This is true for all Coxeter groups.

The affine dihedral subgroups of the affine group $W(A_n)$ can be classified in two classes. They are represented by two different extended diagrams of the dihedral subgroups as shown in Figure 2 where

$$h' = \frac{2h}{h-2} \text{ or } h' = \frac{2h}{h+2} \text{ and } h'' = \frac{2h}{h-1}. \quad (9)$$

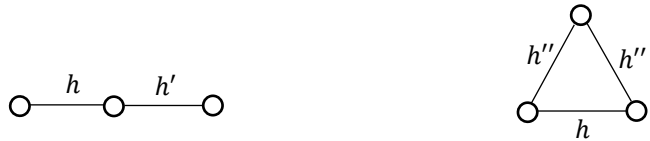

**Figure 2**. Affine dihedral subgroups of the affine groups $W(A_n)$ for odd and even $n$.

Two simple examples are given as the affine dihedral groups one obtained from the affine group $W(A_3)$ and one from $W(A_4)$ as shown in Figure 3. In these examples lattice $A_3$ projects onto a square lattice in the other case the lattice $A_4$ projects onto the Penrose-like quasi lattice with 5-fold symmetry [9].

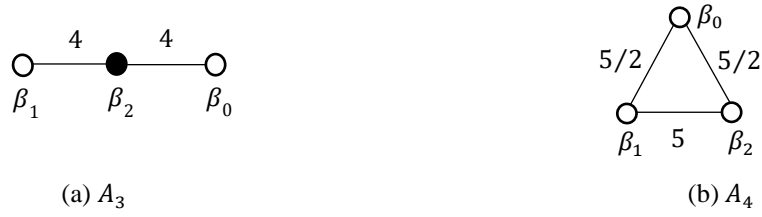

(a) $A_3$   (b) $A_4$

**Figure 3**. Diagrams defining affine dihedral subgroups of affine groups (a) $W(A_3)$ and (b) $W(A_4)$. (Dark node represents a short root)

## 2.2 Affine Subgroup $W(C_n)$ of the affine Coxeter group $W(A_{2n-1})$

Using the simple roots of the diagram $A_{2n-1}$ of Figure 1 we define the following simple roots

$$\alpha'_0 = \alpha_0; \; \alpha'_1 = \frac{\alpha_1 + \alpha_{2n-1}}{2}; \; \alpha'_2 = \frac{\alpha_2 + \alpha_{2n-2}}{2}; \ldots; \; \alpha'_{n-1} = \frac{\alpha_{n-1} + \alpha_{n+1}}{2}; \; \alpha'_n = \alpha_n. \tag{10}$$

They represent the simple roots of the Coxeter- Dynkin diagram $C_n$ where $\alpha'_i$, ($i = 1,2,\ldots,n-1$) are the short roots of $C_n$. Using the simple roots $\alpha'_j$, ($j = 1, 2, \ldots, n$) we can construct the Cartan matrix from the matrix elements $M_{ij} = \frac{2(\alpha'_i, \alpha'_j)}{(\alpha'_j, \alpha'_j)}$ with $\det M = 2$.

With the addition of the root $\alpha'_0$ the Cartan matrix of the extended diagram can be obtained and it can be shown that the determinant of the $(n + 1) \times (n + 1)$ matrix is zero.
The point group $W(c_n)$ is generated by the reflections

$$r'_{\alpha_1} = r_{\alpha_1} r_{\alpha_{2n-1}}, \; r'_{\alpha_2} = r_{\alpha_2} r_{\alpha_{2n-2}}, \ldots, r'_{\alpha_{n-1}} = r_{\alpha_{n-1}} r_{\alpha_{n+1}}, \; r'_{\alpha_n} = r_{\alpha_n}. \tag{11}$$

The reflection generators $r'_{\alpha_1}, r'_{\alpha_2}, \ldots, r'_{\alpha_n}$ generate the point group $W(c_n)$ of order $2^n n!$, a subgroup of the permutation group of order $(2n)!$.
Adding the reflection generator which represents the reflection with respect to the hyperplane bisecting the root $\alpha'_0 = \alpha_0$ usually denoted by the affine reflection $r_{\alpha_0,1}$ the group is extended to the $W(C_n)$ which represents the symmetry of the lattice $D_n$. A particular lattice $D_4$ for $n = 4$ has a particular symmetry represented by quaternions which can be obtained from the lattice $A_7$ which in turn has an octonionic representation [10-11].

## 3. The Group $W(D_n)$

From the simple roots of the diagram $D_n$ we define the simple roots

$$\alpha'_0 = \alpha_0, \; \alpha'_1 = \alpha_1, \ldots, \alpha'_2 = \alpha_{n-2}, \; \alpha'_{n-1} = \frac{1}{2}(\alpha_{n-1} + \alpha_n) \tag{12}$$

and the corresponding generators as

$$r'_{\alpha_{0,1}}, r'_{\alpha_1}, \ldots, r'_{\alpha_{n-1}}. \tag{13}$$

The generator $r'_{\alpha_{0,1}}$ is the affine generator representing reflection with respect to the hyperplane bisecting the root $\alpha'_0$. Then the set of generators of (12) generates the group $W(B_{n-1})$. As a special case take $n = 4$. This shows that $W(B_3)$ is a subgroup of the group $W(D_4)$. The lattice $B_3$ is the simple cubic lattice with the octahedral point symmetry of order 48 which can be embedded in the four-dimensional checkerboard lattice with a point symmetry of order 192 in 4 different ways. In each case, one of the roots of $W(D_4)$ $\alpha_0, \alpha_1, \alpha_2, \alpha_3$ can be taken as an extended root. Actually, the lattice $D_4$ has a larger point symmetry of order $192 \times 6 = 1152$ which is also the symmetry of its Voronoi cell called 24-cell. This is a typical example that a four-dimensional lattice can be projected into the simple cubic lattice.

Now consider the dihedral subgroup of the Coxeter-Weyl group $W(d_n)$ whose generators defined by

$$R_1 = r_{\alpha_1} r_{\alpha_3} \ldots r_{\alpha_{n-1}}, R_2 = r_{\alpha_2} r_{\alpha_4} \ldots r_{\alpha_n} \text{ for } n \text{ even} \qquad (14)$$

$$R_1 = r_{\alpha_1} r_{\alpha_3} \ldots r_{\alpha_n}, R_2 = r_{\alpha_2} r_{\alpha_4} \ldots r_{\alpha_{n-1}} \text{ for } n \text{ odd} \qquad (15)$$

and $n \geq 4$.

The Coxeter number $h = 2n - 2$ is always an even number. In both cases the point dihedral group $W(I_2(h))$ of order $2h$ has the generation relation

$$R_1^2 = R_2^2 = (R_1 R_2)^h = 1. \qquad (16)$$

Adding the generator $R_0 = r_{\alpha_{0,1}}$ and noting that $r_{\alpha_0} = (R_1 R_2)^{h/2}$ then the diagram of the affine dihedral group can be taken as in Figure 4.

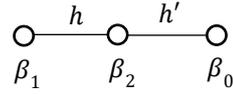

**Figure 4**. Coxeter-Dynkin diagram of the affine dihedral group (see the text for the definition of $h'$).

For $n = 4$ the dihedral group is the group $W(g_2)$ of order 12 and obtained from the diagram $D_4$ by defining the roots $\alpha_0, \alpha_2$ and $\frac{1}{3}(\alpha_1 + \alpha_3 + \alpha_4)$ and the corresponding generators as described earlier [8].

There is one more special group in this category is that the icosahedral group $W(h_3)$ is a non-crystallographic subgroup of $W(d_6)$ [12-13]. Both groups have the same Coxeter number 10. Projection of the root lattice $D_6$ into the 3D Euclidean space describes the icosahedral symmetric quasicrystallography, an important prediction of the Coxeter groups in the condensed matter physics. Generators of $W(H_3)$ can be taken as $R_0 = r_{\alpha_{0,1}}$, $R_1 = r_{\alpha_1} r_{\alpha_5}$, $R_2 = r_{\alpha_2} r_{\alpha_4}$ and $R_3 = r_{\alpha_3} r_{\alpha_6}$ satisfying the generation relation

$$R_0^2 = R_1^2 = R_2^2 = R_3^2 = (R_1 R_2)^3 = (R_2 R_3)^5 = (R_0 R_2)^{5/2} = 1. \qquad (17)$$

In terms of the extended Coxeter-Dynkin diagrams, they are illustrated as in Figure 5.

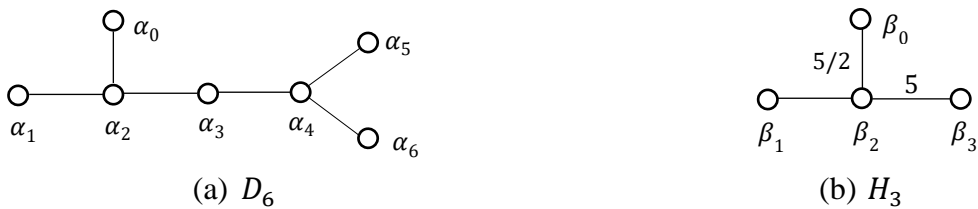

(a) $D_6$      (b) $H_3$

**Figure 5**. Coxeter-Dynkin diagrams of affine $D_6$ and affine $H_3$.

Here the roots of $W(h_3)$ can be obtained from the roots of $W(d_6)$ as follows:

$$\beta_1 = \frac{1}{\sqrt{2(\tau+2)}}(\alpha_1 + \tau\alpha_5), \beta_2 = \frac{1}{\sqrt{2(\tau+2)}}(\alpha_2 + \tau\alpha_4), \beta_3 = \frac{1}{\sqrt{2(\tau+2)}}(\tau\alpha_3 + \alpha_6) \tag{18}$$

and the extended root $\beta_0 = -\tau(\beta_1 + 2\beta_2 + \tau\beta_3)$. The generators in (17) now refers the reflections with respect to the roots $\beta_i$, $(i = 0,1,2,3)$. In the 3D Euclidean space where the roots can be written in terms of orthonormal vectors the extended root $\beta_0$ takes a simpler form $\beta_0 = \sqrt{2}(0,-1,0)$. Hence $R_{0,1}$ represents the reflection with respect to the plane bisecting the vector $\sqrt{2}(0,-1,0)$.

## 4. The Groups $W(E_n)$

We will consider $E_6$, $E_7$ and $E_8$ in turn.

### 4.1 The group $W(E_6)$

There are two point groups $W(I_2(12))$ and the $W(f_4)$ with the Coxeter number 12 which can be embedded in $W(e_6)$. Take the generators $R_1 = r_{\alpha_1}r_{\alpha_3}r_{\alpha_5}, R_2 = r_{\alpha_2}r_{\alpha_4}r_{\alpha_6}$ which satisfy the generation relation $R_1^2 = R_2^2 = (R_1R_2)^{12} = 1$. Defining $R_{0,1} = r_{\alpha_{0,1}}$ and noting that $r_{\alpha_0} = R_2(R_1R_2)^6$ the generation relation of the affine dihedral subgroup of the group $W(E_6)$ is given by the generation relation

$$R_0^2 = R_1^2 = R_2^2 = (R_1R_2)^{12} = (R_{0,1}R_1)^{12/7} = 1 \tag{19}$$

and represented by the diagram of Figure 6.

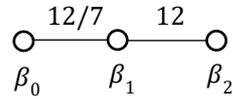

**Figure 6**. Affine Coxeter-Dynkin diagram of the affine dihedral group $W(I_2(12))$ as the subgroup of $W(E_6)$.

The affine Coxeter group $W(F_4)$ is obtained from the affine Coxeter group $W(E_6)$ by defining the roots of $F_4$ in terms of the roots of $E_6$ as follows

$$\alpha'_0 = \alpha_0, \alpha'_1 = \alpha_6, \alpha'_2 = \alpha_3, \alpha'_3 = \frac{1}{2}(\alpha_2 + \alpha_4), \alpha'_4 = \frac{1}{2}(\alpha_1 + \alpha_5). \tag{20}$$

Corresponding reflection generators can be defined as

$$r'_{\alpha_{0,1}} = r_{\alpha_{0,1}}, r'_{\alpha_1} = r_{\alpha_6}, r'_{\alpha_2} = r_{\alpha_3}, r'_{\alpha_3} = r_{\alpha_2}r_{\alpha_4} \text{ and } r'_{\alpha_4} = r_{\alpha_1}r_{\alpha_5}. \tag{21}$$

The diagrammatic representation is illustrated in Figure 7.

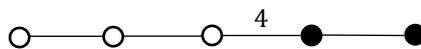

**Figure 7**. Extended Coxeter-Dynkin diagram of $F_4$.

The root lattice $F_4$ is determined by the root lattice $D_4$. Indeed, the root system for $F_4$ consists of first two layers of norms 2 and 4 in the root lattice $D_4$.

## 4.2 The group $W(E_7)$

In this case we have only dihedral subgroup $W(I_2(18))$ having the same Coxeter number with the point Coxeter group $W(e_7)$. The generators of $W(I_2(18))$ can be obtained from the generators of $W(e_7)$ as $R_1 = r_{\alpha_1}r_{\alpha_3}r_{\alpha_5}r_{\alpha_7}, R_2 = r_{\alpha_2}r_{\alpha_4}r_{\alpha_6}$ and adding the affine generator $R_{0,1} = r_{\alpha_{0,1}}$ we obtain the generation relation of the affine extension of the dihedral group $W(I_2(18))$ as

$$R_{0,1}{}^2 = R_1{}^2 = R_2{}^2 = (R_1R_2)^{18} = (R_{0,1}R_2)^{9/5} = 1. \tag{22}$$

## 4.3 The group $W(E_8)$

Construction of the affine dihedral subgroup of the affine group $W(E_8)$ follows the general procedure. Defining the generators $R_{0,1} = r_{\alpha_{0,1}}$, $R_1 = r_{\alpha_1}r_{\alpha_3}r_{\alpha_5}r_{\alpha_7}$, $R_2 = r_{\alpha_2}r_{\alpha_4}r_{\alpha_6}r_{\alpha_8}$. The point dihedral subgroup $W(I_2(30))$ of the point Coxeter group $W(e_8)$ are generated by the generators $R_1$ and $R_2$. Adding to them the affine generator $R_{0,1} = r_{\alpha_{0,1}}$ we obtain the generation relation of the affine extension of the dihedral group $W(I_2(30))$ as

$$R_{0,1}{}^2 = R_1{}^2 = R_2{}^2 = (R_1R_2)^{30} = (R_{0,1}R_1)^{15/7} = 1. \tag{23}$$

The point Coxeter group $W(e_8)$ has another special non crystallographic subgroup $W(h_4)$[14] whose generators can be defined as $R_1 = r_{\alpha_1}r_{\alpha_7}$, $R_2 = r_{\alpha_2}r_{\alpha_6}$, $R_3 = r_{\alpha_3}r_{\alpha_5}$ and $R_4 = r_{\alpha_4}r_{\alpha_8}$ satisfying the generation relation

$$R_1{}^2 = R_2{}^2 = R_3{}^2 = R_3{}^2 = (R_1R_2)^3 = (R_2R_3)^3 = (R_3R_4)^5 = 1. \tag{24}$$

Adding the affine generator $R_{0,1} = r_{\alpha_{0,1}}$ we obtain the generation relation of the affine group $W(H_4)$ as

$$R_{0,1}{}^2 = R_1{}^2 = R_2{}^2 = R_3{}^2 = R_4{}^2 = (R_1R_2)^3 = (R_2R_3)^3 = (R_3R_4)^5 = (R_{0,1}R_1)^{5/2} = 1. \tag{25}$$

The diagrammatic representations are illustrated in Figure 8.

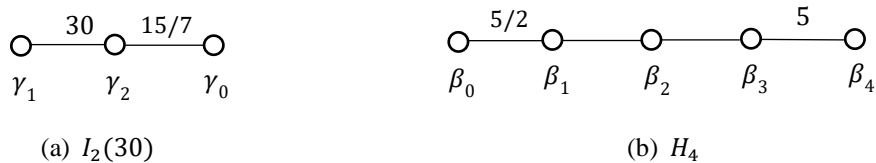

(a) $I_2(30)$          (b) $H_4$

**Figure 8**. Coxeter-Dynkin diagrams of the of the affine dihedral group $W(I_2(30))$ and the affine group $W(H_4)$.

The simple roots of $W(h_4)$ can be written as

$$\beta_1 = \frac{1}{\sqrt{2(\tau+2)}}(\alpha_1 + \tau\alpha_7), \quad \beta_2 = \frac{1}{\sqrt{2(\tau+2)}}(\alpha_2 + \tau\alpha_6),$$
$$\beta_3 = \frac{1}{\sqrt{2(\tau+2)}}(\alpha_3 + \tau\alpha_5), \quad \beta_4 = \frac{1}{\sqrt{2(\tau+2)}}(\tau\alpha_4 + \alpha_8) \tag{26}$$

and the extended root can be written as $\beta_0 = -2\tau\beta_1 - (3\tau + 1)\beta_2 - 2\tau^3\beta_3 - \tau^4\beta_4$.

Generators in (24) correspond to the reflections $\beta_i, (i = 0, 1, 2, 3, 4)$ where $R_{0,1}$ is the reflection with respect to the hyperplane bisecting the root $\beta_0$. In 4D Euclidean space roots $\beta_i$ can be represented in terms of orthonormal vectors where $\beta_0$ can be written as $\beta_0 = \sqrt{2}(0, 0, 0, -1)$ and the generator $R_{0,1}$ represents the reflection with respect to the hyperplane bisecting the root $\beta_0$.

## 5. Conclusion

We have determined the subgroups of the Coxeter groups having the same Coxeter number with the parent Coxeter group. Two classes of subgroups seem to be useful in the quasicrystallographic structures. One class of subgroups is the dihedral subgroups of the Coxeter groups which play an important role in the classification of the planar quasicrystallography. Another very interesting subgroup is the icosahedral group $W(h_3)$ obtained as a subgroup of $W(d_6)$ which represents the point symmetry of the icosahedral symmetric quasicrystallography so that the affine group $W(H_3)$ describe the symmetry of the icosahedral symmetric quasicrystals. The affine subgroup $W(H_4)$ of the affine Coxeter group $W(E_8)$ represents the quasicrystallographic structures in 4D Euclidean space [15] which admits $W(H_3)$ as a subgroup.

## References

1. Conway, J. H., Sloane N. J. A. (1988). Sphere Packings, Lattices and Groups. Springer-Verlag New York Inc.
2. Conway, J. H., Sloane N. J. A. (1991). ed Hilton, P., Hirzebruch, F. F. & Remmert, R. The cell structures of certain lattices. In Miscellanea Mathematica. New York, Springer. pp. 71-108.
3. Humphreys, J. E. (1990). Reflection Groups and Coxeter Groups, Cambridge, Cambridge University Press.
4. Coxeter, H.S. M. (1973). Regular Complex Polytopes, Cambridge, Cambridge University Press.
5. Koca, N. O., Al-Siyabi, A., Koca, M. & Koc, R. (2019). Prototiles and Tilings from Voronoi and Delone Cells of the Root Lattice An, *Symmetry*, 11, 1082.
6. Ziegler, G.M. (1995). *Lectures on Polytopes*, Graduate Texts in Mathematics, 152, Springer-Verlag, New York, Inc.
7. Senechal, M. (1995). Quasicrystals and Geometry, Cambridge, Cambridge University Press.
8. Carter, R. W. (1972). Simple Groups of Lie Type, John Wiley & Sons Ltd.
9. Koca, M., Koca, N. O. & Koc, R. (2014). Affine Coxeter group $Wa(A4)$, quaternions, and decagonal quasicrystals. *Int. J. Geom. Methods Mod. Phys*. **11**(4), 1450031.
10. Karsch, F. and Koca, M. (1990). G$_2$(2) as the automorphism group of the octonionic root system of E$_7$. J. *Phys. A: Math. & Gen*. **A23**, 4739.
11. Koca, M., Koc, R. & Koca, N. O. (2007). The Chevalley group G2 (2) of order 12096 and the octonionic root system of E7. *Linear. Alg. Appl.* **422**, 808.
12. Kramer, P. (1993). Modeling of quasicrystals. *Physica Scripta*, T**49**, 343-348.
13. Koca, N. O., Koc, R., Koca, M. A. Al-Siyabi (2021). Dodecahedral structures with Mosseri–Sadoc tiles. *Acta Cryst*, **A77**, 105-116.
14. Koca, M., Koc, R. & Koca, N. O. (1998). Quaternionic roots of E8 related Coxeter graphs and quasicrystals, Tr. J. of Physics **22 (5)**, Article 8, 421-435.
15. Elser, V., Sloane, N. J. A. (1987). A highly symmetric four-dimensional quasicrystal, *J. Phys. A: Math. & Gen*. 1987, **20**, 6161.